\documentclass[conference]{./sty/IEEEtran}

\ifCLASSINFOpdf
  \usepackage[pdftex]{graphicx}
  \graphicspath{{./fig/}}
\else
  \usepackage[dvips]{graphicx}
  \graphicspath{{./fig/}}
\fi

\usepackage{amssymb}
\usepackage[cmex10]{amsmath}

\DeclareMathAlphabet{\mathitbf}{OML}{cmm}{b}{it}

\newcommand{\ve}{\mathbf}
\newcommand{\m}{\mathbf}
\newcommand{\mf}[1]{\mathbf{\tilde{\mathbf{#1}}}} % matrix in frequency domain
\newcommand{\vef}[1]{\mathbf{\tilde{\mathbf{#1}}}} % vector in frequency domain
\begin{document}
%
% paper title
\title{Coded OFDM by Unique Word Prefix}

% author names and affiliations
% use a multiple column layout for up to three different
% affiliations
\author{
\IEEEauthorblockN{Mario Huemer,~\emph{Senior Member,~IEEE,} and Christian Hofbauer}
\IEEEauthorblockA{Klagenfurt University\\
Institute of Networked and Embedded Systems \\
Universitaetsstr. 65-67, 9020 Klagenfurt\\
mario.huemer@uni-klu.ac.at, chris.hofbauer@uni-klu.ac.at}
\and
\IEEEauthorblockN{Johannes B. Huber,~\emph{Fellow,~IEEE}}
\IEEEauthorblockA{University of Erlangen-Nuremberg\\
Institute for Information Transmission \\
Cauerstr. 7, D-91058 Erlangen \\
huber@lnt.de}}

% make the title area
\maketitle

\begin{abstract}
In this paper we propose a novel transmit signal structure and an adjusted and optimized receiver for OFDM 
(orthogonal frequency division multiplexing). Instead of the conventional cyclic prefix we us a
deterministic sequence, which we call unique word (UW), as guard interval. We show how unique words, which are
already well investigated for single carrier systems with frequency domain equalization (SC/FDE), can also be introduced 
in OFDM symbols. Since unique words represent known sequences, they can advantageously be used for synchronization
and channel estimation purposes. Furthermore, the proposed approach introduces a complex number Reed-Solomon (RS-) code 
structure within the sequence of subcarriers. This allows for RS-decoding or to apply a 
highly efficient Wiener smoother succeeding a zero forcing stage at the receiver. 
We present simulation results in an indoor multipath environment to highlight the advantageous properties of the 
proposed scheme.
\end{abstract}

\IEEEpeerreviewmaketitle

\section{Introduction}
\let\thefootnote\relax\footnotetext{Christian Hofbauer has been funded by the European Regional Development Fund
and the Carinthian Economic Promotion Fund (KWF) under grant 20214/15935/23108.} 

In conventional OFDM signaling, subsequent 
symbols are separated by guard intervals, which are usually implemented as cyclic prefixes (CPs) \cite{VanNee00}.
By this, linear convolution of the signal with the channel impulse response is transformed 
into a cyclic convolution which allows for 
a low complex equalization in frequency domain. In this paper, we propose to use 
known sequences, which we call unique words (UWs), instead of cyclic prefixes. The 
technique of using UWs had already been investigated in-depth for SC/FDE systems, where the 
introduction of unique words in time domain is straight forward \cite{Witschnig02_1}, 
since the data symbols are also defined in time domain. In this paper, we will show 
how unique words can be introduced in OFDM time domain symbols, even though the data QAM (quadrature 
amplitude modulation) symbols are defined in frequency domain. Furthermore, we will introduce an optimized
receiver concept adjusted to the novel transmit signal structure.  

Figure \ref{fig:CP_UW} compares the transmit data structure of CP- and UW-based transmission in time
domain \cite{Huemer10_1}. Both structures make sure that linear convolution of an OFDM symbol with the 
impuls response of a dispersive (e.g. multipath) channel
appears as a cyclic convolution at the receiver side. Nevertheless, there are also
some fundamental differences between CP- and UW-based transmission:

\begin{figure}[!ht]
\centering
\includegraphics[width=3.3in]{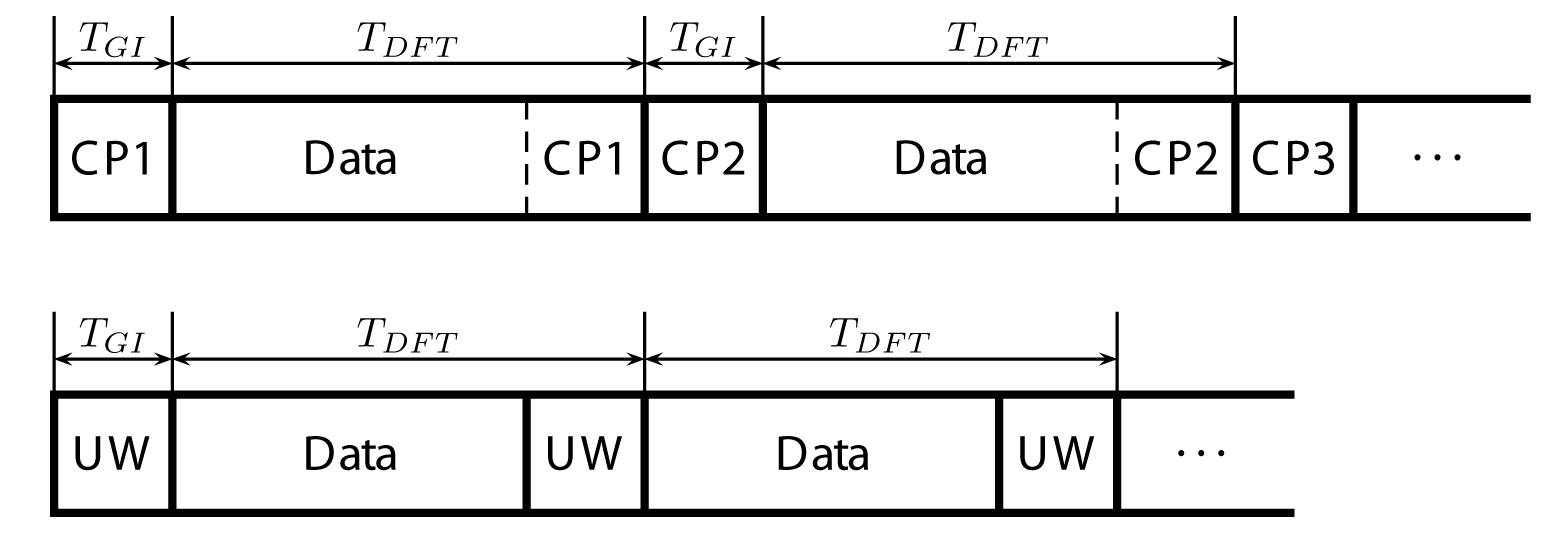}
\caption{Transmit data structure using CPs (above) or UWs (below).}
\label{fig:CP_UW}
\end{figure}

\begin{itemize}
\item The UW is part of the DFT (discrete Fourier transfom)-interval, whereas the CP is not. 
\item The CP is random, whereas the UW is a known deterministic sequence. Therefore the 
UW can advantageously be utilized for synchronization \cite{Huemer03_2} and channel estimation purposes 
\cite{Witschnig03}. 
\end{itemize}

Both statements hold for OFDM- as well as for SC/FDE-systems. However, in OFDM - different 
to SC/FDE - the introduction of UWs in time domain leads to another fundamental and beneficial signal property: 

\begin{itemize}
\item A UW in time domain generates a word of a complex number RS-code (cf. e. g. \cite{Redinbo00}) or of a 
specific coset to such a code in frequency domain, i.e along the subcarriers. Therefore, the UW
could be exploited for error correction or (more appropriate) for erasure correction for highly
attenuated subcarriers. Another interpretation of this fact which we prefer here, is an introduction
of correlations along the subcarriers. These correlations can advantageously be used as a-priory knowledge
at the receiver to significantly improve the BER (bit error ratio) behavior in frequency selective 
environments.
\end{itemize}

\smallskip
The rest of the paper is organized as follows: In section II we describe our approach of how to introduce unique words in 
OFDM symbols. Section III introduces an LMMSE (linear minimum mean squared
error) receiver that exploits the a-priory
knowledge introduced at the transmitter side. In section IV the novel UW-OFDM concept is compared to the
classical CP-OFDM by means of simulation results. For this, the IEEE 802.11a WLAN (wireless local area networks) 
standard serves as reference system. \smallskip

\noindent
\textit{Notation} 

Lower-case bold face variables ($\ve{a},\ve{b}$,...) indicate vectors, and upper-case bold face variables 
($\m{A},\m{B}$,...) indicate matrices. To distinguish between time and frequency domain variables, we use a tilde
to express frequency domain vectors and matrices ($\vef{a},\mf{A}$,...), respectively. We further use $\mathbb{C}$ to denote the set 
of complex numbers, $\m{I}$ to denote the identity matrix, $(\cdot)^T$ to denote transposition, $(\cdot)^H$ to denote 
conjugate transposition, and $E[\cdot]$ to denote
expectation.

\section{Generation of Unique Words in OFDM Symbols}
In conventional CP-OFDM, the data vector $\vef{d}\in\mathbb{C}^{N_d \times 1}$ is defined in 
frequency domain. Typically, zero subcarriers are inserted at the band edges and
at the DC-subcarrier position, which can mathematically be described by a matrix operation 
\begin{equation}
		\vef{x} = \m{B} \vef{d} 
\end{equation}
with $\vef{x}\in\mathbb{C}^{N \times 1}$ and $\m{B}\in\mathbb{C}^{N \times N_d}$. $\m{B}$ consists of zero-rows 
at the positions of the zero-subcarriers, and of appropriate unit row vectors at the positions of data-subcarriers.
The vector $\vef{x}$ denotes the OFDM symbol in frequency domain. The vector of time domain samples 
$\ve{x}\in\mathbb{C}^{N \times 1}$ is calculated via an IDFT 
(inverse DFT) operation, which can conveniently be formulated 
in matrix notation by $\ve{x}=\m{F}_N^{-1} \vef{x}$. Here, $\m{F}_N$ is the $N$-point-DFT
matrix defined by $\m{F}_N=\left(F_{mn}\right)$ with 
$F_{mn}=w^{mn}$ for $m=0,1,...,N-1$, $n=0,1,...,N-1$, and with $w=\mathrm{e}^{-j2\pi/N}$.  

\begin{figure}[!ht]
\centering
\includegraphics[width=2.0in]{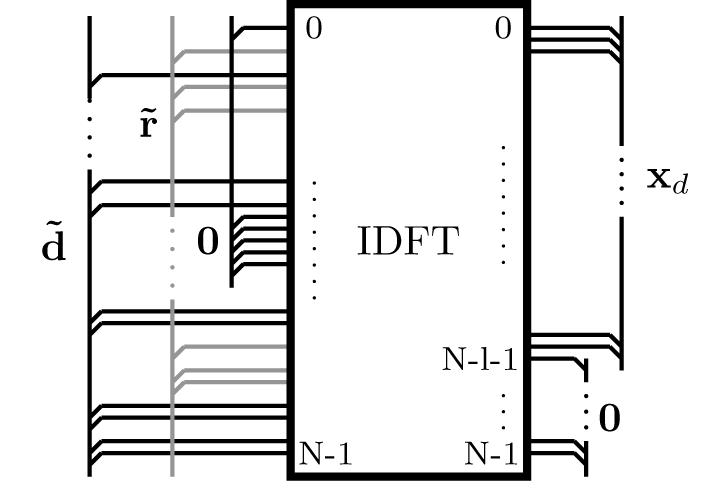}
\caption{Time- and frequency-domain view of an OFDM symbol in UW-OFDM.}
\label{fig:uw_ofdm_overview}
\end{figure}

\smallskip
We now modify this conventional approach by introducing a pre-defined sequence 
$\ve{x}_u$ with $\ve{x}_u\in\mathbb{C}^{l \times 1}$, which we call unique word, and which shall 
form the tail of the time domain vector, which we now denote by  $\ve{x}'$. Hence, $\ve{x}'$ consists of two parts and is given 
by $\ve{x}' = \begin{bmatrix}\ve{x}_d^T&\ve{x}_u^T\end{bmatrix}^T$, where 
$\ve{x}_d\in\mathbb{C}^{(N-l) \times 1}$ and $\ve{x}_u\in\mathbb{C}^{l \times 1}$. 
The vector $\ve{x}_u$ represents the UW of length $l$, and thus only $\ve{x}_d$ is random and affected 
by the data. In order to simplify subsequent descriptions, but w.l.o.g. we use a two-step approach for the so-defined vector $\ve{x}'$	: 
\begin{itemize}
\item In a first step we will generate a zero UW
	\begin{equation}
		\ve{x} = \begin{bmatrix}\ve{x}_d \\ \ve{0}\end{bmatrix},
	\end{equation}
	such that $\ve{x}=\m{F}_N^{-1} \vef{x}$.
\item In a second step we will determine the transmit symbol by 
	\begin{equation}
			\ve{x}' = \ve{x} + \begin{bmatrix}\ve{0} \\ \ve{x}_u\end{bmatrix}. \label{equ:6}
	\end{equation}
\end{itemize}
We now describe the first step in detail: As in conventional OFDM, the QAM data symbols and the 
zero-subcarriers are specified in frequency domain in vector $\vef{x}$, but here in addition the zero-word
is specified in time domain as part of the 
vector $\ve{x}$. As a consequence, the linear system of equations
\begin{equation} 
		\ve{x}=\m{F}_N^{-1} \vef{x}	\label{equ:1}
\end{equation}
can only be fulfilled by reducing the number $N_d$ of data 
subcarriers, and by introducing a set of redundant subcarriers instead. We let 
the redundant subcarriers form the vector $\vef{r}\in\mathbb{C}^{l \times 1}$, 
further introduce a permutation matrix $\m{P}\in\mathbb{C}^{(N_d+l) \times (N_d+l)}$, 
and form an OFDM symbol (containing $N-N_d-l$ zero-subcarriers) in frequency domain by 
\begin{equation}
		\vef{x} = \m{B} \m{P} \begin{bmatrix} \vef{d} \\ \vef{r} \end{bmatrix}. \label{equ:2}
\end{equation}
We will detail the reason for the introduction of the permution matrix and its specific construction shortly below.
Figure \ref{fig:uw_ofdm_overview} illustrates this approach in a graphical way: The input of the IDFT block is composed of 
data subcarriers ($\vef{d}$), zero subcarriers, and redundant subcarriers ($\vef{r}$), which are distributed 
over the entire non-zero part of vector $\vef{x}$ as specified by the permutation matrix $\m{P}$. The output of the IDFT block,
which corresponds to the vector $\ve{x}$ of time domain samples of an OFDM symbol, is composed of the random 
part $\ve{x}_d$, and the zero UW $\ve{0}$. 

By inserting equation (\ref{equ:2}) into (\ref{equ:1}), the relation between the time and the 
frequency domain representation of the OFDM symbol can be written as
\begin{equation}
		\m{F}_N^{-1} \m{B} \m{P} \begin{bmatrix} \vef{d} \\ \vef{r} \end{bmatrix} = \begin{bmatrix}\ve{x}_d \\ \ve{0} \end{bmatrix}.
\end{equation}
With
\begin{equation}
		\m{M}=\m{F}_N^{-1} \m{B} \m{P}= \begin{bmatrix} \m{M}_{11} & \m{M}_{12} \\ \m{M}_{21} & \m{M}_{22}\end{bmatrix}, 
\end{equation}
where $\m{M}_{ij}$ are appropriate sized sub-matrices, it follows that $\m{M}_{21} \vef{d} + \m{M}_{22} \vef{r} = \ve{0}$, and hence 
$\vef{r} = -\m{M}_{22}^{-1}\m{M}_{21} \vef{d}$.
With the matrix
\begin{equation}
		\m{T} = -\m{M}_{22}^{-1}\m{M}_{21} \label{equ:30}
\end{equation}
($\m{T}\in\mathbb{C}^{l \times N_d}$), the vector of redundant subcarriers can thus be determined by the linear mapping
\begin{equation}
		\vef{r} = \m{T} \vef{d}, \label{equ:3}
\end{equation}
which (despite of the permutation) exactly corresponds to a complex number RS-code
construction along the subcarriers (i.e. $l$ subsequent zeros in the transform domain). We summarize, 
that the IDFT of a frequency domain vector as given in equation (\ref{equ:2}), which is composed of a data part 
$\vef{d}$, a set of zero-subcarriers, and a part $\vef{r}$ of redundant subcarriers, which is determined by equation (\ref{equ:3}), 
results in a time domain vector, which features the zero UW $\ve{0}$ at its tail. 

We notice that the construction of $\m{T}$, and therefore also the variances of the redundant subcarriers highly depend on the 
positions of the redundant subcarriers within the entire frequency domain vector
$\vef{x}$. Hence, the permutation matrix $\m{P}$ 
has to be chosen carefully. We select $\m{P}$ such that $\mathrm{trace}\left(\m{T}\m{T}^H\right)$
becomes minimum \cite{Huemer10_2}. This provides minimum energy on the redundant subcarriers on average (when averaging over all possible data 
vectors $\vef{d}$). In section \ref{sec:simulation_results} we will specify the permutation matrix $\m{P}$ for our simulated 
system setup. 

Note that equation (\ref{equ:3}) introduces correlation in the vector $\vef{x}$ of frequency 
domain samples of an OFDM symbol. This can advantageously be utilized in an
optimized receiver structure as it will be shown 
in the next section. In the following we use the notation $\vef{s}$ with
\begin{equation}
		\vef{s} = \m{P} \begin{bmatrix} \vef{d} \\ \vef{r} \end{bmatrix} 
						= \m{P} \begin{bmatrix} \m{I} \\ \m{T} \end{bmatrix} \vef{d}
						= \m{U} \vef{d}, \label{equ:31}
\end{equation}
($\vef{s}\in\mathbb{C}^{(N_d+l) \times 1},\m{U}\in\mathbb{C}^{(N_d+l) \times N_d}$) for the non-zero part of $\vef{x}$, such that
$\vef{x} = \m{B} \vef{s}$. \smallskip

In the second step the transmit symbol $\ve{x}'$ is generated by adding the unique word as described in equation (\ref{equ:6}). 
The frequency domain version $\vef{x}_u \in\mathbb{C}^{N \times 1}$ of the UW is defined by $\vef{x}_u = \m{F}_N \begin{bmatrix} \ve{0}^T & \ve{x}_u^T \end{bmatrix}^T$. Note that $\ve{x}'$ can also be written as $\ve{x}'= \m{F}_N^{-1} (\vef{x}_u + \vef{x}) = \m{F}_N^{-1} (\vef{x}_u + \m{B} \vef{s}$).

\section{LMMSE UW-OFDM Receiver}
At the receiver side the UW may be exploited for error and/or erasure correction as usual for RS-codes. 
Here, we restrict the explanations to an exploitation of correlations between subcarriers which proofed to
be more appropriate, since the redundant subcarrier symbols - unlike the data symbols - usually do not fit 
to a discrete grid (e.g. odd Gaussian integers). After the transmission over a multipath channel and after the common 
DFT operation, the non-zero part $\vef{y}\in\mathbb{C}^{(N_d+l) \times 1}$ of a received OFDM frequency domain symbol 
can be modeled as
\begin{equation}
		\vef{y} = \m{B}^T \m{F}_N \m{H} \m{F}_N^{-1} (\vef{x}_u + \m{B} \vef{s}) + \m{B}^T \m{F}_N \ve{n},
\end{equation}
where $\m{H}$ denotes a cyclic convolution matrix with $\m{H}\in\mathbb{C}^{N\times N}$, and $\ve{n}\in\mathbb{C}^{N \times 1}$ 
represents a noise vector with the covariance matrix $\sigma_n^2 \m{I}$. The multiplication with $\m{B}^T$ excludes 
the zero subcarriers from further operation. The matrix $\m{F}_N \m{H} \m{F}_N^{-1}$ is diagonal and contains the sampled channel frequency response on its main diagonal. $\mf{H} = \m{B}^T \m{F}_N \m{H} \m{F}_N^{-1} \m{B}$ with $\mf{H}\in\mathbb{C}^{(N_d+l) \times (N_d+l)}$ is 
a down-sized version of the latter excluding the entries corresponding to the zero-subcarriers. The received symbol can therefore 
also be written as
\begin{equation}
		\vef{y} = \mf{H} (\m{B}^T \vef{x}_u + \vef{s}) + \m{B}^T \m{F}_N \ve{n}.
\end{equation}
As usual for conventional OFDM, we propose to apply a zero forcing equalization by multiplying with $\mf{H}^{-1}$ from the left. 
This results in 
\begin{equation}
		\vef{y}^{'} = \mf{H}^{-1} \vef{y}
						= \m{B}^T \vef{x}_u + \vef{s} + \vef{v}
\end{equation}
with the noise vector $\vef{v} = \mf{H}^{-1} \m{B}^T \m{F}_N \ve{n}$. After its usage for synchronization and/or 
channel estimation purposes, the UW can be extracted from further operation by	
\begin{equation}
		\vef{y}^{''} = \vef{y}^{'} - \m{B}^T \vef{x}_u = \vef{s} + \vef{v}.
\end{equation}
We mention, that the influence of the UW could also already be eliminated earlier, by simply subtracting 
$\mf{H} \m{B}^T \vef{x}_u$ from $\vef{y}$. 

$\vef{s}$ contains the data as well as the redundant subcarrier symbols. Since the redundant subcarrier symbols 
have been calculated out of the data symbols by equation (\ref{equ:3}), they are correlated with the data symbols 
and among each other. Because of that we propose to apply an LMMSE Wiener smoother \cite{Kay93} on $\vef{y}^{''}$, which results in the 
noise reduced estimate
\begin{equation}
		\widehat{\vef{s}} = \m{C}_{\tilde{s}\tilde{s}} \left(\m{C}_{\tilde{s}\tilde{s}} + \m{C}_{\tilde{v}\tilde{v}} \right)^{-1} \vef{y}^{''},
\end{equation}
where $\m{C}_{\tilde{s}\tilde{s}},\m{C}_{\tilde{v}\tilde{v}}\in\mathbb{C}^{(N_d+l) \times (N_d+l)}$ denote the covariance matrices of $\vef{s}$ and $\vef{v}$, respectively.
Let us take a closer look on these covariance matrices: 
\begin{eqnarray}
		\m{C}_{\tilde{s}\tilde{s}} 	&=& E \left[\vef{s} \vef{s}^H \right] \nonumber \\
																&=& E \left[(\m{U}\vef{d}) (\m{U}\vef{d})^H \right] \nonumber \\
																&=& E \left[\m{U}\vef{d} \vef{d}^H \m{U}^H \right] \nonumber \\
																&=& \m{U} E \left[\vef{d} \vef{d}^H \right] \m{U}^H. 
\end{eqnarray}
Assuming uncorrelated and zero-mean data QAM symbols with variance
$\sigma_d^2$, we obtain a constant matrix
\begin{equation}
		\m{C}_{\tilde{s}\tilde{s}} = \sigma_d^2 \m{U} \m{U}^H,
\end{equation}
that has to be calculated only once and can be determined in advance, cf. equations (\ref{equ:30}), (\ref{equ:31}). For $\m{C}_{\tilde{v}\tilde{v}}$ we have
\begin{eqnarray}
		\m{C}_{\tilde{v}\tilde{v}} 	&=& E \left[\vef{v} \vef{v}^H \right]  \nonumber \\
																&=& E \left[(\mf{H}^{-1}\m{B}^T\m{F}_N\vef{n}) (\mf{H}^{-1}\m{B}^T\m{F}_N\vef{n})^H \right] \nonumber \\
																&=& E \left[\mf{H}^{-1}\m{B}^T\m{F}_N\vef{n}\vef{n}^H \m{F}_N^H\m{B}(\mf{H}^{-1})^H\right] \nonumber \\
																&=& \mf{H}^{-1}\m{B}^T\m{F}_N E \left[\vef{n}\vef{n}^H \right] \m{F}_N^H\m{B}(\mf{H}^{-1})^H \nonumber \\
																&=& \sigma_n^2 \mf{H}^{-1}\m{B}^T\m{F}_N \m{F}_N^H\m{B}(\mf{H}^{-1})^H \nonumber \\
																&=& N \sigma_n^2 \mf{H}^{-1}\m{B}^T\m{B}(\mf{H}^{-1})^H \nonumber \\
																&=& N \sigma_n^2 \mf{H}^{-1}(\mf{H}^{-1})^H. \label{equ:11}
\end{eqnarray}
Let $H(f_i)$ with $i=0,1,...,N_d+l-1$ denote the channel frequency response at the corresponding subcarrier frequency $f_i$. Now equation
(\ref{equ:11}) can also be written as 
\begin{equation}
		\m{C}_{\tilde{v}\tilde{v}} = N \sigma_n^2 \mathrm{diag}\left\{\frac{1}{\left|H(f_0)\right|^2},\frac{1}{\left|H(f_1)\right|^2},...\right\}.
\end{equation}
$\m{C}_{\tilde{v}\tilde{v}}$ depends on the noise variance $\sigma_n^2$ and on the channel frequency response. With the Wiener smoothing
matrix
\begin{equation}
		\mf{W} = \m{C}_{\tilde{s}\tilde{s}} \left( \m{C}_{\tilde{s}\tilde{s}} +  \m{C}_{\tilde{v}\tilde{v}} \right)^{-1}
\end{equation}
the operations performed on a received OFDM frequency domain symbol $\vef{y}$ can now be compactly written as
\begin{equation}
			\widehat{\vef{s}} = \mf{W} \mf{H}^{-1} (\vef{y}-\mf{H} \m{B}^T \vef{x}_u).
\end{equation}

\begin{figure*}[ht]
\centering
\includegraphics[width=6in]{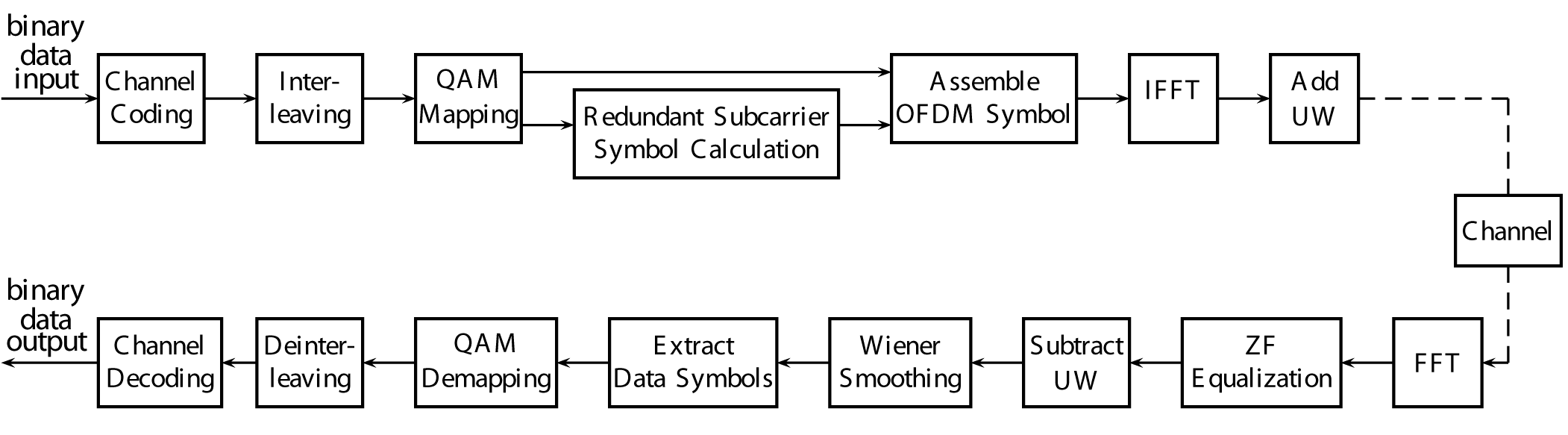}
\caption{Block diagram for simulation analysis.}
\label{fig:block_diagram_ISIT}
\end{figure*}	

\noindent Finally, the data part $\widehat{\vef{d}} = \begin{bmatrix} \m{I} & \m{0} \end{bmatrix} \m{P}^{-1} \widehat{\vef{s}}$
can be processed further as usual. We notice, that the error $\vef{e}=\vef{s}-\widehat{\vef{s}}$ has zero mean, 
and its covariance matrix is given by $\m{C}_{\tilde{e}\tilde{e}}= \left(\m{I} - \mf{W}\right) \m{C}_{\tilde{s}\tilde{s}}$ \cite{Kay93}.
$\m{C}_{\tilde{e}\tilde{e}}$ can further be used in the case when additional channel coding is 
applied. Especially, varying noise variance along the subcarriers within the data vector $\vef{d}$ may be exploited 
as well known from coded transmission over time variant channels, cf. e.g. \cite{Huber84}.\smallskip	
		
We summarize the receiver operations per OFDM symbol:
\begin{itemize}
\item Perform a DFT operation to obtain $\vef{y}$.
\item Eliminate the influence of the UW by subtracting $\mf{H} \m{B}^T \vef{x}_u$ from $\vef{y}$.
\item Apply ZF-equalization followed by an LMMSE Wiener smoothing operation. (Of course, 
both operations can be implemented in one combined single matrix multiplication operation.)
\item Extract the data part and process it further as usual. 
\end{itemize}
Note again, that the ZF equalization and the UW elimination can be exchanged as described above.

\section{Simulation Results} \label{sec:simulation_results}
Figure \ref{fig:block_diagram_ISIT} shows the block diagram of the simulated UW-OFDM system (equivalent complex baseband 
description is used throughout this paper). After channel coding, 
interleaving and QAM-mapping, the redundant subcarrier symbols are determined using equation (\ref{equ:3}). 
After assembling the OFDM symbol, which is composed of $\vef{d}$, $\vef{r}$, and a set of zero-subcarriers, 
the IFFT (inverse fast Fourier transform) is performed. Finally the UW is added in time domain. At the receiver the
FFT (fast Fourier transform) operation is followed by a ZF equalization as in classical CP-OFDM. Next the frequency 
domain version of the UW is subtracted. Then the Wiener smoother is applied to the symbol, and finally demapping, 
deinterleaving and decoding is performed. For the soft decision Viterbi decoder the main diagonal of matrix $\m{C}_{\tilde{e}\tilde{e}}$ 
is used to specify the varying noise variances along the subcarriers after equalization and Wiener filtering. 

We compare our novel UW-OFDM approach with the classical CP-OFDM concept. The
IEEE 802.11a WLAN standard \cite{IEEE99} serves as
reference system. We apply the same parameters for UW-OFDM as in \cite{IEEE99} wherever possible: $N=64$,
sampling frequency $f_s = 20$MHz, DFT period $T_{DFT}=3.2\mu s$, guard duration
$T_{GI}=800ns$. Instead of 48 data subcarriers and 4 pilots we use $N_d=36$
data subcarriers and $l=16$ redundant subcarriers. 
The zero subcarriers are chosen as in \cite{IEEE99}, the indices of the redundant subcarriers are chosen to be
\{2, 6, 10, 14, 17, 21, 24, 26, 38, 40, 43, 47, 50, 54, 58, 62\}. This choice, which can easily also be 
described by equation (\ref{equ:2}) with an appropriately constructed matrix $\m{P}$, minimizes the 
total energy of the redundant subcarriers on average (when averaging over all possible data 
vectors $\vef{d}$) \cite{Huemer10_2}.

In our approach the unique
word shall take over the synchronization tasks which are normally performed
with the help of the 4 pilot subcarriers. In order to make a fair comparison,
the energy of the UW related to the total energy of a transmit symbol is set to 4/52, which exactly corresponds
to the total energy of the 4 pilots related to the total energy of a transmit symbol in the IEEE standard. 
Note that in conventional CP-OFDM like in the WLAN standard, the total length of an OFDM symbol is given by $T_{GI}+T_{DFT}$. 
However, the guard interval is part of the DFT period in our approach. Therefore, both systems show comparable bandwidth efficiency. 

\begin{figure}[!ht]
\centering
\includegraphics[width=3.3in]{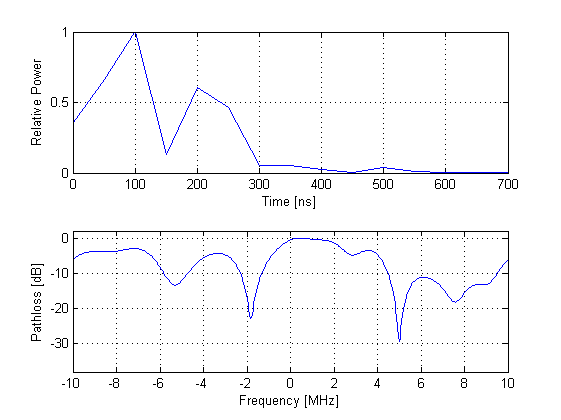}
\caption{Time- and frequency-domain representation of the used multipath channel snapshot.}
\label{fig:channel}
\end{figure}

The multipath channel has been modeled as a tapped delay line, each tap with uniformly distributed
phase and Rayleigh distributed magnitude, and with power decaying exponentially. A detailed description of
the model can be found in \cite{Witschnig03}. Figure \ref{fig:channel} shows one typical channel snapshot featuring an rms 
delay spread of 100ns. The frequency response shows two spectral 
notches within the system's bandwidth.

\begin{figure}[!ht]
\centering
\includegraphics[width=3.5in]{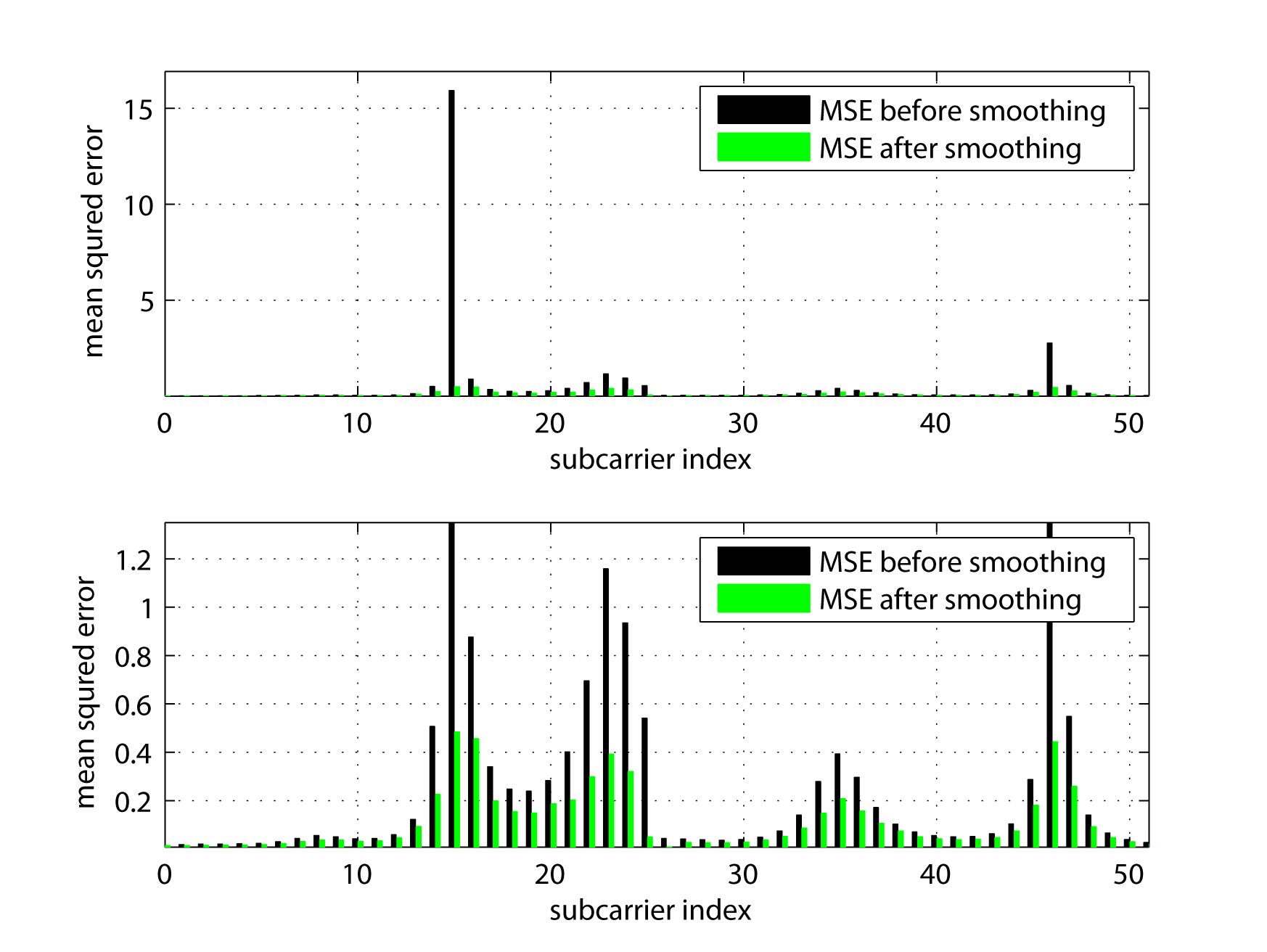}
\caption{Noise reduction effect of the Wiener smoother in a frequency selective environment for $E_b/N_0$ = 15dB. Above: full scale; below:
zoomed y-axis.}
\label{fig:Smoothing_effect}
\end{figure}

In order to clearly demonstrate the effect of our UW-OFDM approach with the derived LMMSE receiver, the following 
discussions are based on results obtained for the displayed channel snapshot. Figure \ref{fig:Smoothing_effect} 
compares the mean squared errors on the $N_d+l$ (data + redundant) subcarriers before and after the Wiener smoothing 
operation. We note that all subcarriers experience a significant noise reduction by the smoother, but the effect is 
impressive on the subcarriers corresponding to spectral notches in the channel frequency response. The subcarriers 
with index 15 and 46 correspond to the spectral notches around 5MHz and -2MHz, respectively, cf. figure \ref{fig:channel}. 

\begin{figure}[!ht]
\centering
\includegraphics[width=3.3in]{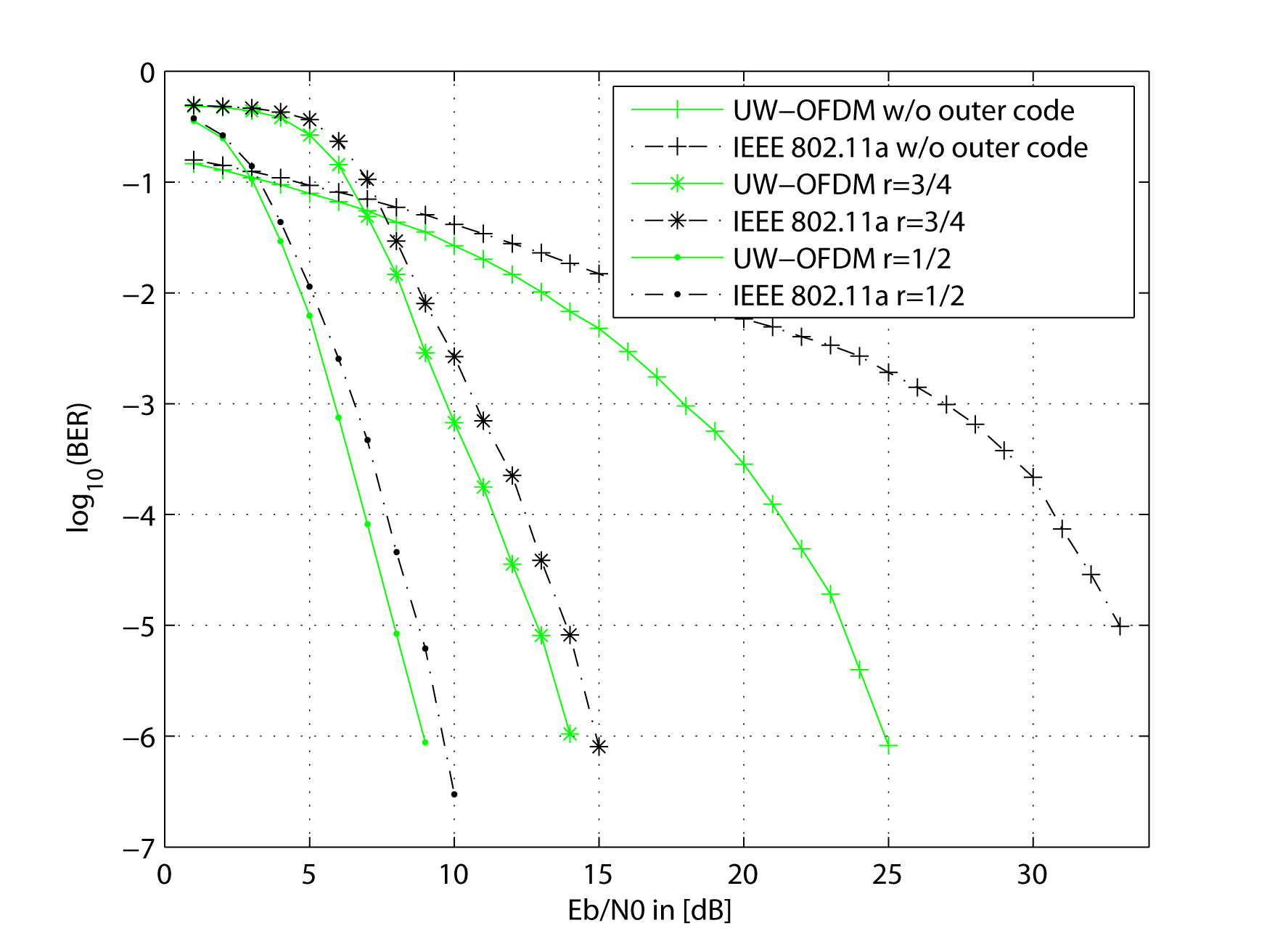}
\caption{BER comparison between the novel UW-OFDM approach and the IEEE 802.11a standard for the channel snapshot displayed above.}
\label{fig:BER}
\end{figure}

In figure \ref{fig:BER} the BER-behavior of the IEEE 802.11a standard and the novel UW-OFDM approach are compared, 
both in QPSK-mode for the channel displayed in figure \ref{fig:channel}. The channel snaphot represents a typical 
indoor NLOS (non line of sight) office environment, nevertheless further simulation results for different delay spreads 
and for time varying channels are in preparation for upcoming publications. Here we show results of simulations 
with and without the usage of an additional outer code. The outer code features the coding rates $r=3/4$ and $r=1/2$, 
respectively. Both systems use the same convolutional coder with the industry standard rate 1/2, constraint length 7 
code with generator polynomials (133,171). For $r=3/4$ puncturing 
is used as described in \cite{IEEE99}. Note that due to the different number of data symbols per OFDM symbol, 
the interleaver had to be slightly adapted compared to the WLAN standard. Perfect channel knowlegde is assumed in 
both approaches. In the case of no further outer code, the gain achieved by the LMMSE smoother is impressive. 
This can be explained by the significant noise reduction on heavily attenuated subcarriers or by the fact, that UW-OFDM 
includes a complex number RS-code in a natural way and channel coding provides high gains especially for varying channels. 
For the coding rates $r=3/4$ and $r=1/2$ the novel UW-OFDM approach still achieves a gain of 0.9dB and 0.65dB at 
a bit error ratio of $10^{-6}$, respectively. 

\section{Conclusion}
In this work we introduced a novel OFDM signaling concept, where the guard intervals are built by unique words instead
of cyclic prefixes. The proposed approach introduces a complex number Reed-Solomon code 
structure within the sequence of subcarriers. As an important conclusion we can state, that besides the possibility to use the UW for synchronization and channel estimation purposes, the novel approach additionally allows to apply a highly efficient LMMSE Wiener smoother, 
which significantly reduces the noise on the subcarriers, especially on highly attenuated subcarriers. 
Simulation results illustrate, that the novel approach outperforms classical CP-OFDM in a typical frequency selective indoor scenario.

\end{document}